\documentclass[seceq]{ptptex}
\usepackage{wrapft}

\usepackage{graphicx}
\usepackage{bm}

\bmdefine{\bk}{k}
\bmdefine{\bx}{x}
\bmdefine{\br}{r}

\newcommand{\HII}{H{\sc ii} }

\newcommand{\OVII}{O{\sc vii} }

\newcommand{\simgt}{\lower.5ex\hbox{$\; \buildrel > \over \sim \;$}}
\newcommand{\simlt}{\lower.5ex\hbox{$\; \buildrel < \over \sim \;$}}

\pubinfo{Vol.~158, August 2005}

\title{
Cosmic Structure Formation at Low and High Redshifts
}

\author{
Naoki \textsc{Yoshida}$^{1,}$\footnote{E-mail: nyoshida@a.phys.nagoya-u.ac.jp} 
}

\inst{
$^1$Department of Physics, Nagoya University, Nagoya, Aichi 464-8602, Japan
}

\recdate{
January 31, 2005}

\abst{
The currently standard theory of cosmic structure formation 
posits that the present-day clumpy appearance of the universe developed 
through gravitational amplification of the matter density fluctuations
that are generated in the very early universe. 
The energy content of the univese and
the basic statistics of the initial density field 
have been determined with a reasonable 
accuracy from recent observations of the cosmic microwave background,
large-scale structure, and distant supernovae. 
It has become possible to make accurate predictions 
from the standard model.
Cosmology is now at the stage where we can rigourously test
the model against various observations of large- and small-scale 
structure. We review the latest observations and the recent progress
in the theory of structure formation at low and high redshifts. 
Two promising methods to probe large-scale 
matter distribution are introduced and the future prospects are discussed. 
Results from state-of-the-art cosmological simulations are also presented. 
}

\begin{document}
\maketitle

\section{Introduction}
Observations of extragalactic objects suggest that the universe is 
approximately homogeneous and isotropic at large scales. 
The almost perfectly isotropic feature in the cosmic background 
radiation temperature also manifests that the universe {\it was} 
homogeneous and isotropic, while a variety of clumpy 
structures are seen in the local universe, such as galaxies and 
galaxy clusters. 
One also finds, for instance in the catalogues of galaxy redshift 
surveys, that there are some patterns or prominent ``structures'' 
which extend over tens of mega-parsecs (see Fig.~1).
Recent observations of high-redshift galaxies revealed that 
large-scale structure already existed at $z=4-6$, when the age of the 
Universe was just one tenth of the present age.\cite{rf:Shimasaku,rf:Ouchi}
Apparently the universe has undergone a rapid transition from a smooth 
initial state to the clumpy state as we see today, but details remain 
largely unknown. Understanding the origin and evolution of the 
structure of the universe is hence a major goal in modern cosmology. 

The so-called standard theory of structure formation posits that the 
present-day clumpy appearance
of the universe developed through gravitational amplification 
of the initial matter density fluctuations together with other physical
processes. 
This basic picture is now supported by an array of observations,
including the measurement of the cosmic microwave background 
anisotropies by the WMAP satellite.\cite{rf:Spergel03}
The WMAP observation also confirmed that the density fluctuations in the 
early universe arise from adiabatic perturbations 
whose statistics are described by a Gaussian field,\cite{rf:Komatsu} 
as predicted by popular inflationary theories.  
The cosmological parameters that describes 
the dynamics of the universe as a whole 
are now known to a good accuracy. 
Given these bases, and aided by detailed computer simulations,
theoretical models are now able to make accurate predictions
for a variety of properties from galaxy clustering to 
gravitational lensing statistics.

 Over the past two decades, cosmological models based on cold dark 
matter (CDM) have been the most successful, and 
a variant of the CDM model that invokes dark energy has emerged as 
the current leading model.
While there are still some unsolved issues and possible conflictions
with observations,\cite{rf:Primack}
the model is now accepted as providing the basic framework
of cosmic structure formation.

\begin{figure}
  \centerline{\includegraphics[width=14.5cm]{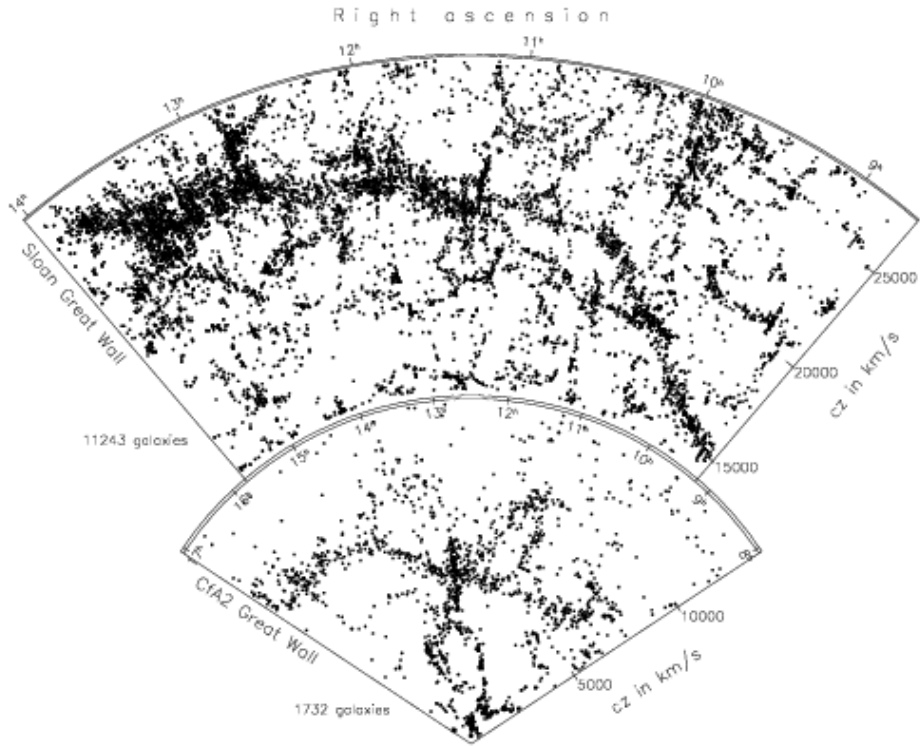}}
  \caption{The large-scale structure in the local universe.
From Gott et al. (2003)\cite{rf:Gott}}
  \label{fig:gal}
\end{figure}

 In this paper, we review recent progress and future 
prospects in the study of structure formation in the universe.
Since a number of excellent articles and reviews are available
on large-scale structure as probed by galaxy redshift surveys,
we refrain from covering the topic. Instead, we introduce
two observational probes that enable to map large-scale 
structure in the distribution of dark matter and that of the 
intergalactic medium.
Large-scale scale structure at high redshift has been recently 
discovered\cite{rf:Shimasaku,rf:Ouchi} and has attracted much attention. 
Observational issues on the ``primeval'' structure will be extensively 
reviewed in the contribution of Okamura in this volume.
Along with these latest observations, we present the results
from state-of-the-art numerical simulations of various kinds.

\section{Probing large-scale structure of the universe}

The distribution of galaxies in the sky has been 
often used for studies on large-scale structure.
\cite{rf:Lick,rf:APM} 
Prominent clustering features were already found 
in the projected galaxy distribution in Lick Catalogue
compiled in 60's.\cite{rf:Lick}
Galaxy redshift surveys added the third dimension, in terms of redshift,
by which one can make a full three-dimensional map of 
the galaxy distribution.\cite{rf:CfA} 
Statistical methods such as two-point correlation functions
\cite{rf:Totsuji,rf:Peebles} and power spectrum
are most often used to quantify the clustering of galaxies,
against which predictions from theoretical models are tested. 
Nowadays these basic statistics
are used to determine cosmological parameters.
The two current-generation redshift surveys, the 2-degree Field 
Survey\cite{rf:2dF} and the Sloan Digital Sky Survey\cite{rf:SDSS}, 
are providing unprecedented data in both quality and quantity. 
Important global quantities and statistics such as luminosity functions,
clustering strengths, and also the properties of galaxy clusters, 
can be obtained with a tremendous precision from the large set of data.

 The best motivation for the CDM model
is its predictive power on the formation of large-scale structure. 
However, extremely large-scale structure ($> 100$ Mpc)
is rarely formed in the CDM model.
Fig.~\ref{fig:gal} shows the largest scale structure found in the
CfA survey and that in the Sloan survey\cite{rf:Gott}.
The CfA ``Great Wall'' extends $\sim 200$ Mpc and the 
Sloan great wall extends nearly twice longer.
Existence of such largest-scale structure may challenge the 
standard model, if commonly discovered in the local 
and distant universe.\cite{rf:Yoshida01}
It remains to be seen whether or
not even larger scale structure exists in our universe.

While the distribution of galaxies provides an overall picture
of matter distribution, there is always a complex issue of ``bias''. 
One usually assumes that galaxies are fair tracer of underlying mass,
introducing a convenient factor called bias.
Estimating bias with respect to the underlying mass
is non-trivial, however. Bias could (or rather, is likely to) depend on
length scale and time, and could be nonlinear with respect to the local 
density.
Hence it would be ideal if one can directly 
map the matter distribution. It can indeed be done
by means of gravitational lensing observations. In the next section, 
we review the recent progress in observations of gravitational
lensing and also its future prospects.

There is a new, recently proposed way to
probe the {\it baryonic} matter distribution in the local universe.
High-level ions of heavy elements such as carbon, nitrogen and oxygen 
in the hot intergalactic medium (IGM) emit photons typically
in soft-Xray bands.
It is expected that next generation X-ray missions can detect 
these emission lines. Since a large fraction of baryons is thought to 
be in such warm/hot phases at the present epoch, future soft-Xray missions
may reveal the location of `missing baryons'. 
Using the redshift of individual metal lines, 
it will be possible to {\it map} the distribution of the hot 
IGM in the near future.
Probing the distribution of dark matter and the diffuse baryonic matter,
together with the galaxy distribution, 
should provide invaluable informations on the process of structure 
formation.

\clearpage

\begin{figure}
  \centerline{\includegraphics[width=14cm]{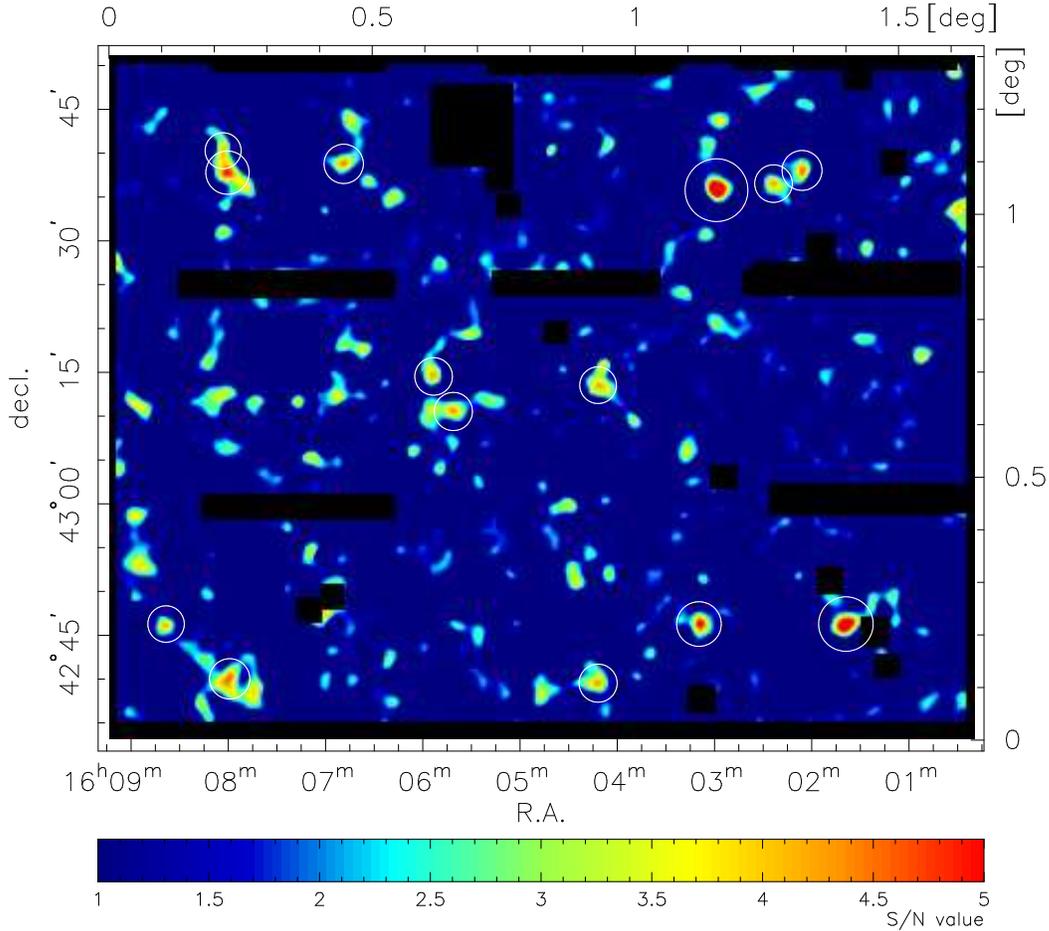}}
  \caption{The weak lensing mass map obtained by
the Subaru Suprime33 pilot survey. The area covers a 2.1 deg$^2$ field,
in which 14 significant peaks (cluster candidates) are found. From Miyazaki,
Hamana, et al. (2002)\cite{rf:GTO}}
  \label{fig:lens}
\end{figure}

\subsection{Dark matter distribution}

Gravitational lensing provides a unique, powerful method
to map the distribution of dark matter.
Weak-lensing technique exploits the deformation
of background galaxies' shapes to map the {\it mass distribution}
in and around large-scale structure.
A number of weak-lensing surveys have been already carried out
and larger-area surveys are being conducted. 
Statistics of the matter distribution can be used to determine the 
cosmological parameters. Indeed, the next generation weak-lensing surveys
are expected to provide the most precise measurements of the matter power 
spectrum. It has recently been proved to be feasible to search for clusters 
of galaxies directly as density enhancements using weak gravitational lensing.
A great advantage of weak lensing is that a 
constructed sample is not biased toward luminous systems which 
optical or X-ray selected catalogs suffer from. 

Fig.~\ref{fig:lens} shows
the weak lensing mass map obtained by the Subaru Suprime33 GTO survey.\cite{rf:GTO}
Fourteen high peaks with $S/N > 4$ are found, among which 11 peaks are
confirmed to be galaxy clusters by follow-up optical observations.
Seven peaks are newly discovered clusters, demonstrating the ability
of finding clusters (massive halos) by weak lensing surveys.
In principle, the halo number counts can be directly comparable
with accurate model predictions\cite{rf:Jenkins,rf:ShethTormen} 
based on the results from $N$-body simulations, and thus can
be used to put strong constraints on cosmological parameters.
While the number of samples in the Subaru GTO survey is still poor, 
the halo number count is consistent with the prediction from the 
$\Lambda$CDM model.

\begin{figure}
  \centerline{\includegraphics[width=7cm,height=6cm]{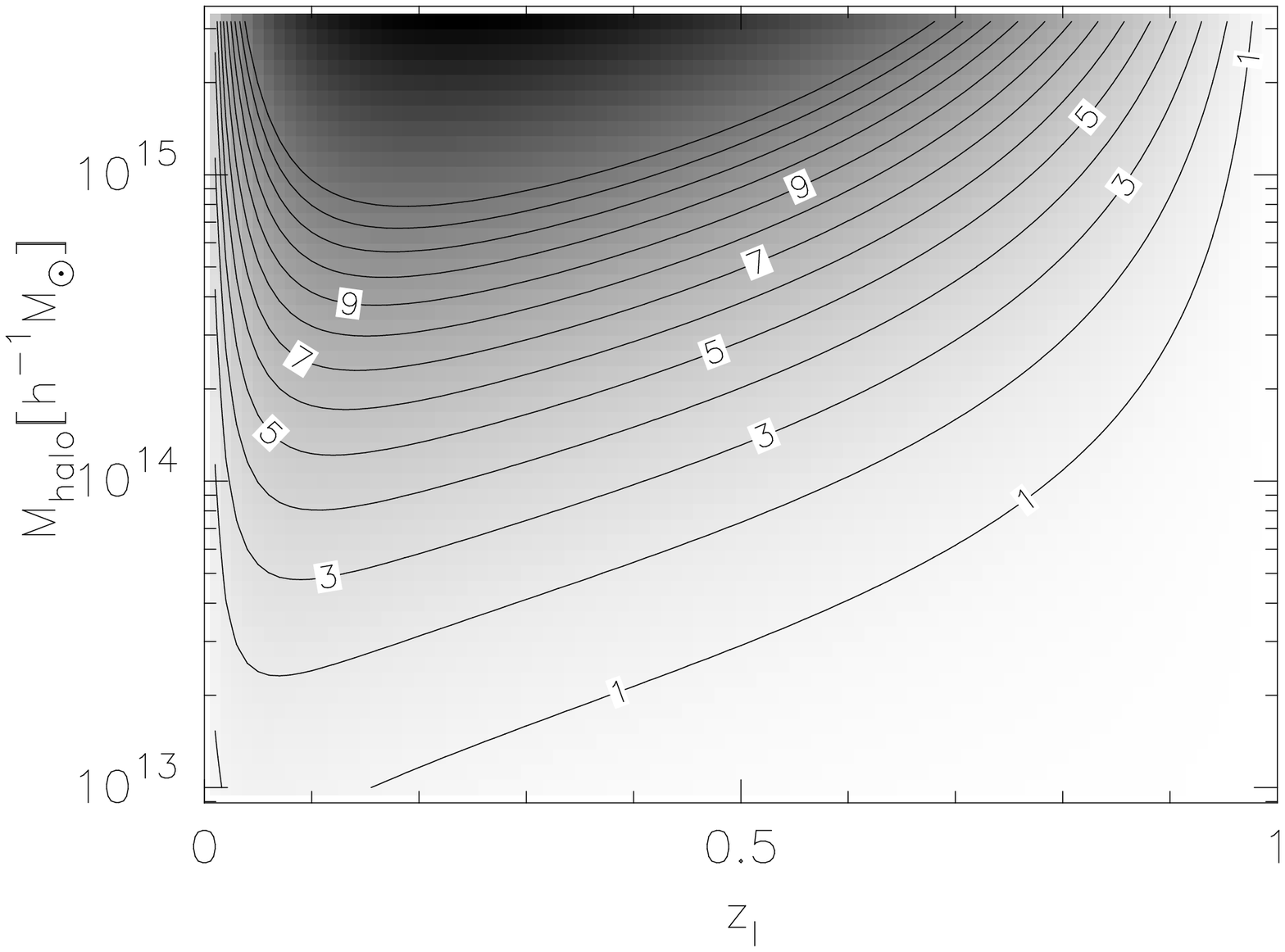}
  \includegraphics[width=7cm,height=7cm]{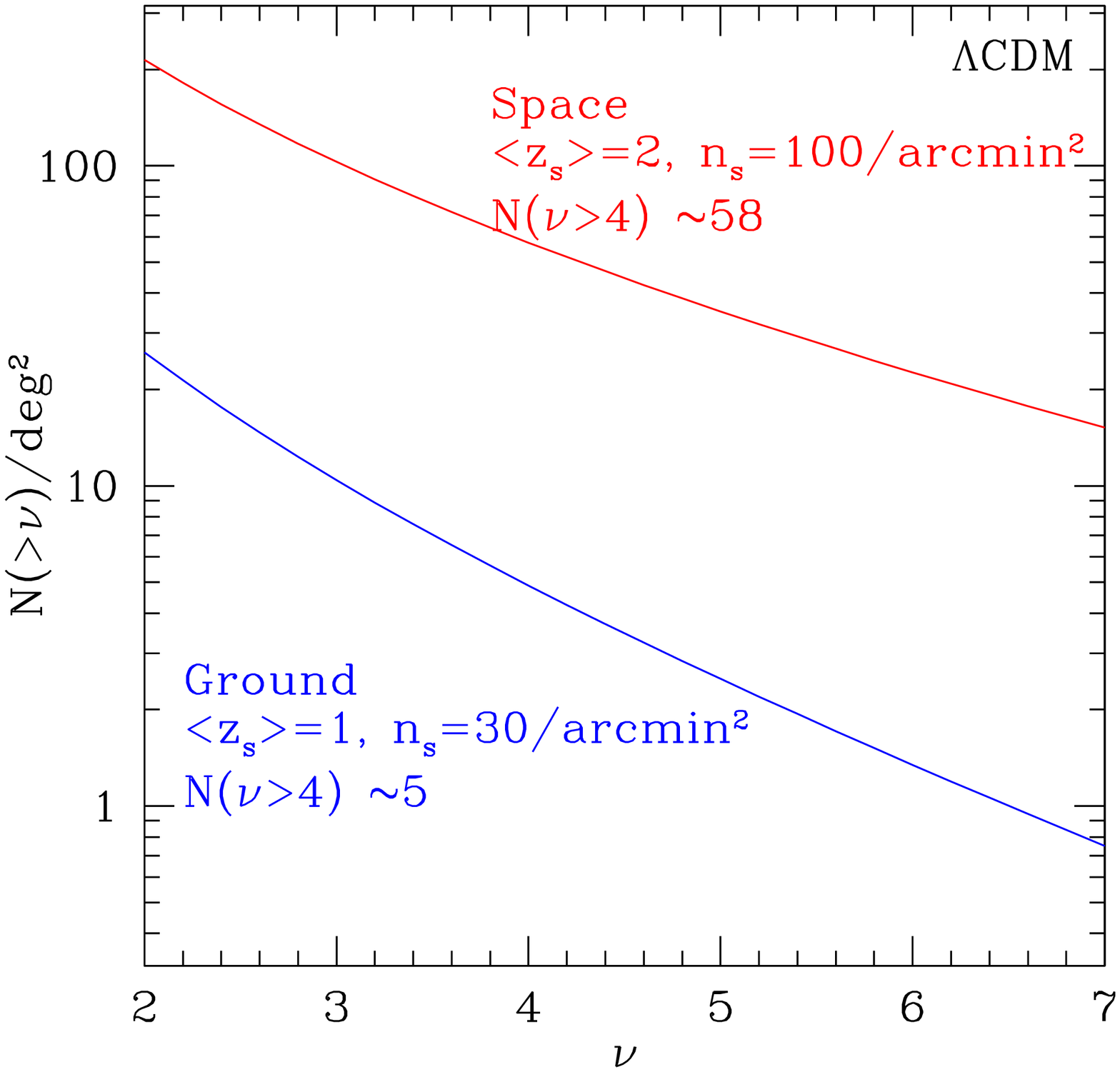}}
  \caption{(Left) The gray scale with contour lines shows the $S/N$ value 
for weak lensing halo detection as a function of halo mass and redshift.
(Right) Predicted number counts 
of peaks which can be detected by a ground-based telescope
and by a space telescope.
From Hamana, Takada \& Yoshida (2004)\cite{rf:HTY04}}
  \label{fig:lens_pred}
\end{figure}

Detailed studies of weak-lensing cluster surveys 
using large $N$-body simulations have been 
recently carried out.\cite{rf:HTY04} 
Fig.~\ref{fig:lens_pred} shows the predicted number counts of halos
with $S/N > \nu$, where $\nu$ is the peak height in the convergence map, 
by two representative observational facilities, a space telescope and a ground-based 
one. The ability of weak-lensing surveys to locate massive dark halos is 
promising. Even with the current generation telescopes, 
the halo detection efficiency is comparable to that of X-ray cluster
search. Future lensing surveys of clusters exploiting a space telescope will
detect $\sim 50-100$ halos with $S/N > 4$. With this high signal-to-noise ratio,
contamination by noise is expected to be small.\cite{rf:HTY04}

\begin{figure}
  \centerline{\includegraphics[width=10cm]{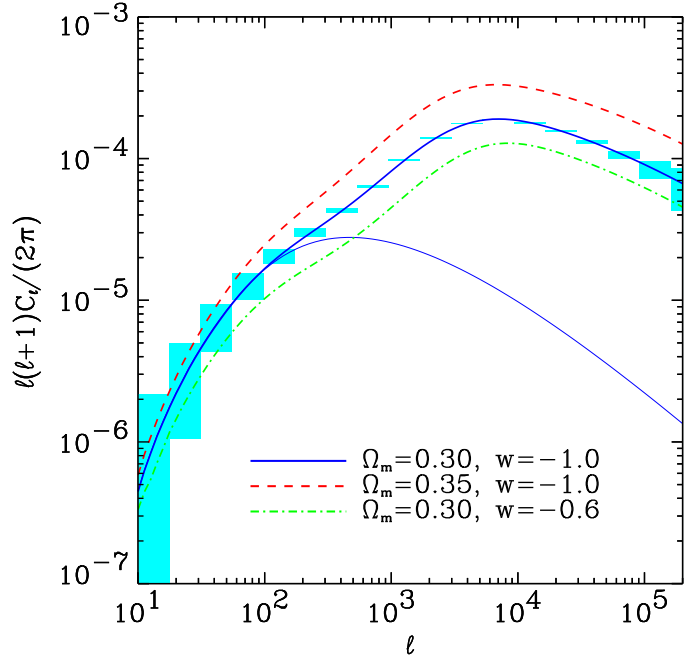}}
  \caption{Prospects for the measurement of the weak-lensing power
spectrum with future surveys. The thick lines are the lensing power spectra
for the indicated three cosmological models. The boxes are estimated
1$\sigma$ errors for the SNAP wide survey. The thin line
is the linear power spectrum. From Refregier (2003).\cite{rf:Refregier}}
  \label{fig:lens_pros}
\end{figure}

 Future weak-lensing surveys will also allow very accurate determination
of cosmological parameters through the lensing power spectrum.
Fig.~\ref{fig:lens_pros} shows the expected accuracy of the 
power spectrum measurement by the SNAP mission.\cite{rf:Refregier}
It is clearly seen that models with different cosmological
parameters can be distinguished. It is also worth noting that
the lensing power spectrum is a sensitive probe of the equation of state 
of dark energy.

\subsection{Missing baryons in the local universe}

The distribution of baryonic matter in the universe remains one of the 
puzzling 
issues. At the present epoch, the total amount of baryons inferred from 
observations of H{\sc i} absorption, gas and stars in galaxies, 
and X-ray emission from hot gas in galaxy clusters is far smaller than 
that predicted by nucleosynthesis calculations\cite{rf:Persic92,rf:FHP98} 
and that determined by measurements of the cosmic microwave background
radiation.\cite{rf:Spergel03}  
Hence, it is now widely believed that about 30-50\% of the baryons in the 
local universe is in yet unknown, dark state.

\begin{figure}
  \centerline{\includegraphics[width=14cm]{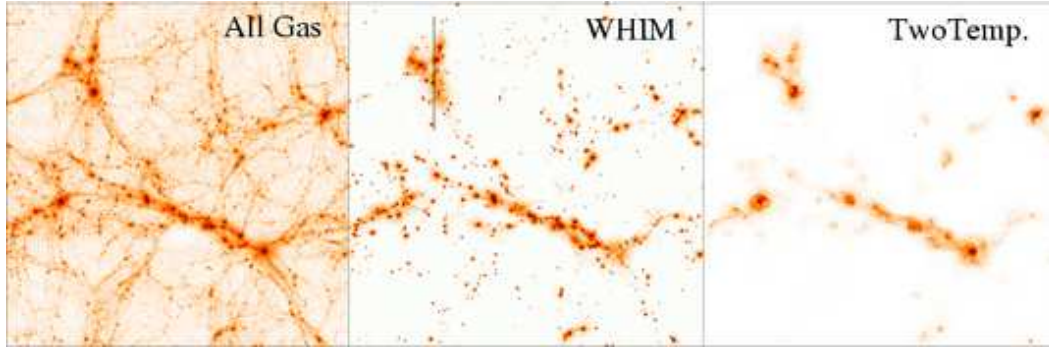}}
  \caption{The distribution of gas (top),
the warm/hot component with $10^6 < T < 10^7$ K (middle),
and the gas that has a two-temperature structure with 
$T_{\rm e}< 0.5T_{\rm i}$ (bottom) in a slab of
$100\times 100\times 20$ ($h^{-1}$Mpc)$^3$. 
From Yoshida, Furlanetto \& Hernquist (2005)\cite{rf:YoshidaWhim}}
\label{fig:whim}
\end{figure}

Numerical simulations of structure formation consistently suggest that 
a large fraction of such {\it missing baryon} is in the warm/hot state 
with temperature $10^5-10^7$K.\cite{rf:CO99,rf:Dave01}
In those simulations, such component is 
found around massive clusters and in filamentary structure.
Fig.~\ref{fig:whim} shows the result from a recent large cosmological 
simulation.\cite{rf:YoshidaWhim}
The warm/hot component is clearly seen as filamentary
structures bridging cluster (high density) regions.
The gas is mostly shock-heated to a temperature of $\sim 10^{5}-10^{7}$ K
during large scale structure formation, and this relatively low
temperature of the gas makes it hard to detect its thermal emission by
conventional X-ray observations.

\begin{figure}
  \centerline{\includegraphics[width=8cm]{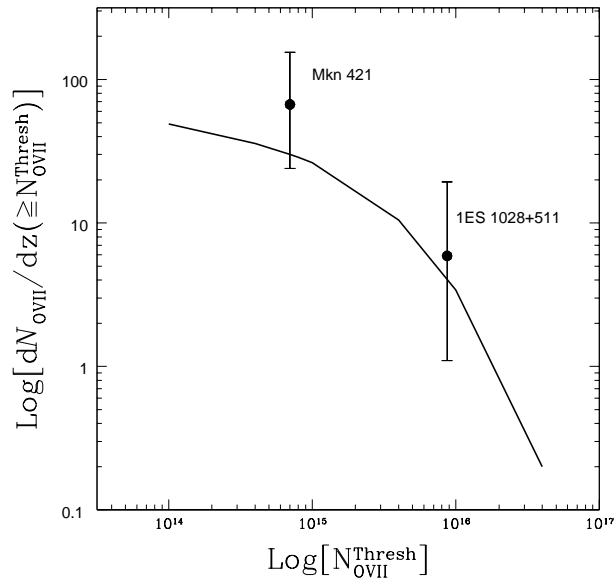}}
  \caption{Predicted (solid line) and observed (points) 
number of \OVII absorbers per unit redshift. From Nicastro et al. (2005)
\cite{rf:Nicastro05}}
\label{fig:whimfrac}
\end{figure}

A variety of observational approaches have been suggested to study the
warm/hot intergalactic medium (WHIM) using either hydrogen or various metal ions.  
With respect to the latter possibility, there
have been several tentative claims that the WHIM has been detected
locally in absorption\cite{rf:Nicastro03} and in emission.\cite{rf:Finoguenov}
Nicastro et al.\cite{rf:Nicastro05} recently estimated the total amount of baryons 
in the warm/hot phase using \OVII absorbers in the spectra of
two blazers. Fig.~\ref{fig:whimfrac} shows the observed number of 
\OVII absorbers per unit redshift, 
compared with the result from a numerical simulation for the $\Lambda$CDM model.\cite{rf:Fang}
They claim that the inferred total mass-density in the warm absorber is
consistent with the theoretical prediction, i.e., roughly half of baryons
in the local universe is in the WHIM.

\begin{figure}
  \centerline{\includegraphics[width=14cm]{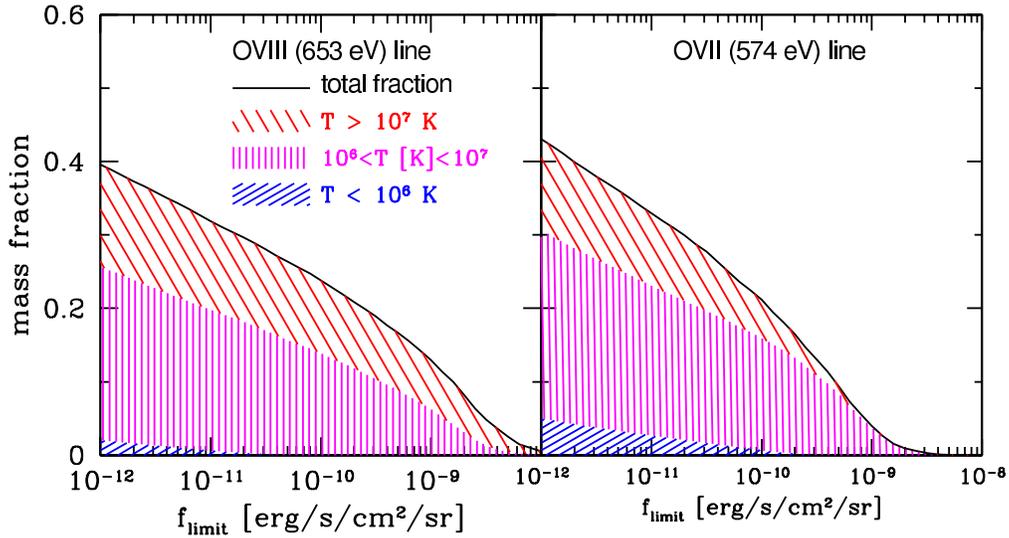}}
  \caption{Estimated baryon mass fraction which can be detected 
through oxygen emission lines. Contributions of baryons with 
$T<10^6 {\rm K}, 10^6<T<10^7 {\rm K},T>10^7 {\rm K}$ are shown separately.
From Yoshikawa et al. (2004)\cite{rf:Yoshikawa05}}
\label{fig:whimfracdios}
\end{figure}
 
Detecting the WHIM in quasar absorption lines (or those of other X-ray sources) 
is always hampered by the fact that observations are limited along line-of-sights 
toward such objects.
Yoshikawa et al.\cite{rf:DIOS} have proposed using O{\sc vii}/O{\sc viii} 
{\it emission} lines to probe the WHIM in
detail with high spectral resolution X-ray detectors.  
From planned configuration and sensitivity of the 
{\it Diffuse Intergalactic Oxygen Surveyor} (DIOS) mission,
they estimate that about half of the WHIM (in mass) can be detected via oxygen 
line emission.
With DIOS, it is possible not only to detect a large fraction of the WHIM 
but also to map the distribution in the local universe because 
redshift of individual lines can be used to determine the distance to the WHIM.
Fig.~\ref{fig:whimfracdios} shows the ability of the proposed 
DIOS mission. A substantial fraction of the hot IGM ($T> 10^6$ K)
can be probed.
These estimates, however, rely on some crucial assumptions
on the IGM metallicity and the relative population of ionization 
levels of oxygen. 

Since the population of different ionization stages 
and the excitation rate of each ion are primarily 
determined by electron impact, it is important to
model appropriately the evolution of {\it electron temperature}
in the WHIM.
In the particular simulation shown in Fig.~\ref{fig:whim}, 
the evolution of electron/ion temperatures
and the relaxation processes are explicitly followed.
It is clearly seen that a bulk of the WHIM has a well-developed
two temperature structure where the electron temperature
is substantially smaller than the ion temperature.
The two-temperature structure of the WHIM has many important
implications. A factor of two systematic shift in temperature, 
typical of the offsets between $\bar{T}$ and $T_{\rm e}$
near shocks, can lead to significant over/under-estimates of 
the ion abundances. Interestingly, when the deviation of the electron 
temperature is taken into account, the emissivity of \OVII lines 
around massive clusters {\it increases}.\cite{rf:YoshidaWhim}
This is because the gas in outskirts of clusters is recently
shock-heated, having a low electron temperature of $\sim$ a few million
degrees, where the \OVII emissivity has a peak.  
The intensity increase (relative to a single temperature model)
can be locally by an order of magnitude, making it promising 
to probe the outer-part of clusters via oxygen emission 
lines.\cite{rf:YoshidaWhim}

\begin{figure}
  \centerline{\includegraphics[width=12cm]{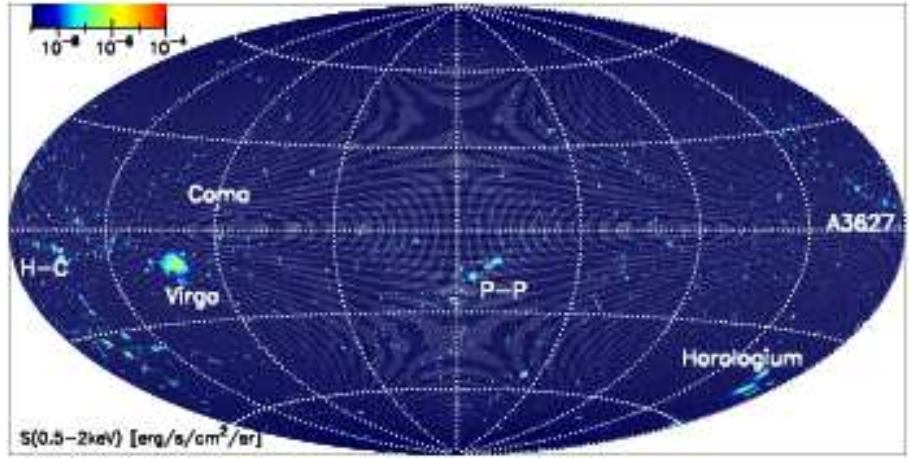}}
  \caption{Distribution of the WHIM in the local universe simulation
of Yoshikawa et al. (2004)\cite{rf:Yoshikawa05} \ The color scale shows the
intensity in soft-Xray band (0.5-2 keV). 
Several large clusters are indicated in the map.}
  \label{fig:whimlocal}
\end{figure}

In the local Universe, there are some prominent large-scale structure
such as Virgo cluster and Pisces-Perseus supercluster region. 
Thus there may be a large amount of WHIM  
also in our local Universe.\cite{rf:Hoffman,rf:Yoshikawa05}
Fig.~\ref{fig:whimlocal} shows the distribution of the WHIM
in the simulated local universe. The simulation is a `constrained
realization' of the local universe starting from a smoothed
linear density field which matches that derived from the IRAS 1.2 Jy 
galaxy survey.\cite{rf:Mathis,rf:IRAS} 
Large-scale structures are seen at approximately right positions with
right sizes.  
These nearby massive clusters will be a primary target of the
next generation soft-Xray mission. It will be a land-marking 
work in observational cosmology if a large fraction of the
{\it missing baryon} (which should then be the majority of baryons 
in the local universe) is finally discovered in our `neighbour'.

\clearpage

\begin{figure}
  \centerline{\includegraphics[width=8cm]{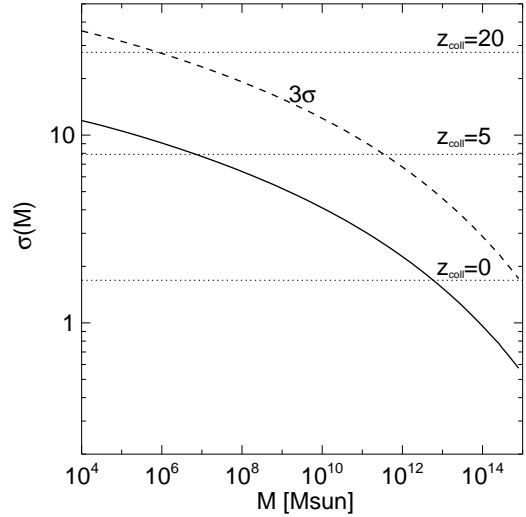}}
  \caption{Mass variance and collapse thresholds in the $\Lambda$CDM model.}
  \label{fig:sigma}
\end{figure}

\section{Hierarchical structure formation and the cosmological first objects}
In this section, we first describe generic features of structure formation
in CDM models. We then discuss the characteristic mass scale of the
cosmological first objects, showing the results from simple analytic models
and those from a detailed numerical simulation.

 The primordial density fluctuations predicted by popular inflation models
have simple characteristics. They are described by a random Gaussian
field, and have a nearly scale-invariant power spectrum $P(k)\propto k^n$ with
$n\sim 1$. Subsequent growth of perturbations in
the radiation-dominated and then in the matter-dominated era
results in a modified power spectrum, but the final shape is still
simple and monotonic in CDM models; the power spectrum 
has a feature that it has progressively larger amplitudes
on smaller length scales. Hence structure formation is expected
to proceed in a ``bottom-up'' manner, with smaller objects forming earlier.

It is useful to work with a properly defined mass variance
to obtain the essence of non-linear evolution and collapse in 
the CDM model.
The mass variance is defined as 
\begin{equation}
\sigma^2 (M)=\frac{1}{2\pi^2}\int P(k) W^2(kR) k^2 {\rm d}k,
\label{eq:variance}
\end{equation}
where the top-hat window function is given by 
$W(x)=3(\sin (x)/x^3 - \cos (x)/x^2)$. 
We also define the threshold over-density for collapse
at $z$ as
\begin{equation}
\delta_{\rm crit} (z)=1.686/D(z),
\end{equation}
where $D(z)$ is the linear growth factor to $z$.
Fig.~\ref{fig:sigma} show the variance and the collapse threshold at
$z=0,5,20$. At $z=20$, the mass of the halos which correspond to 3-$\sigma$ 
fluctuation is just about $10^6 M_{\odot}$.
As shown later in \S\ref{sec:early}, this mass scale coincides with
the characteristic mass of the first objects in which the primordial 
gas can cool and condense. 

It is worth noting that the mass variance is sensitive to the
initial power spectrum. In warm dark matter models in which 
the power spectrum has an exponential cut-off at the particle
free-streaming scale,
\cite{rf:YoshidaWarm}
or in models in which the primordial power spectrum has a `running' 
feature,\cite{rf:Kosowsky,rf:Peiris}
the corresponding mass variance at small mass scales is reduced.
\cite{rf:YoshidaRSI,rf:Somerville}
In such models, early structure formation is effectively delayed, 
and hence nonlinear objects form later than in the CDM model.
Thus the formation epoch of the first objects and hence that of 
cosmic reionization have a direct link to 
the nature of dark matter and the primordial density fluctuations.

There are excellent reviews available on the study of
the formation of the first stars and reionization of the universe.
\cite{rf:BL04,rf:Ciardi,rf:Glover}, so we do not attempt to cover a 
broad range of the topics here. Rather, we focus on a few important issues
in the study of early structure formation.

\begin{figure}
  \centerline{\includegraphics[width=12cm]{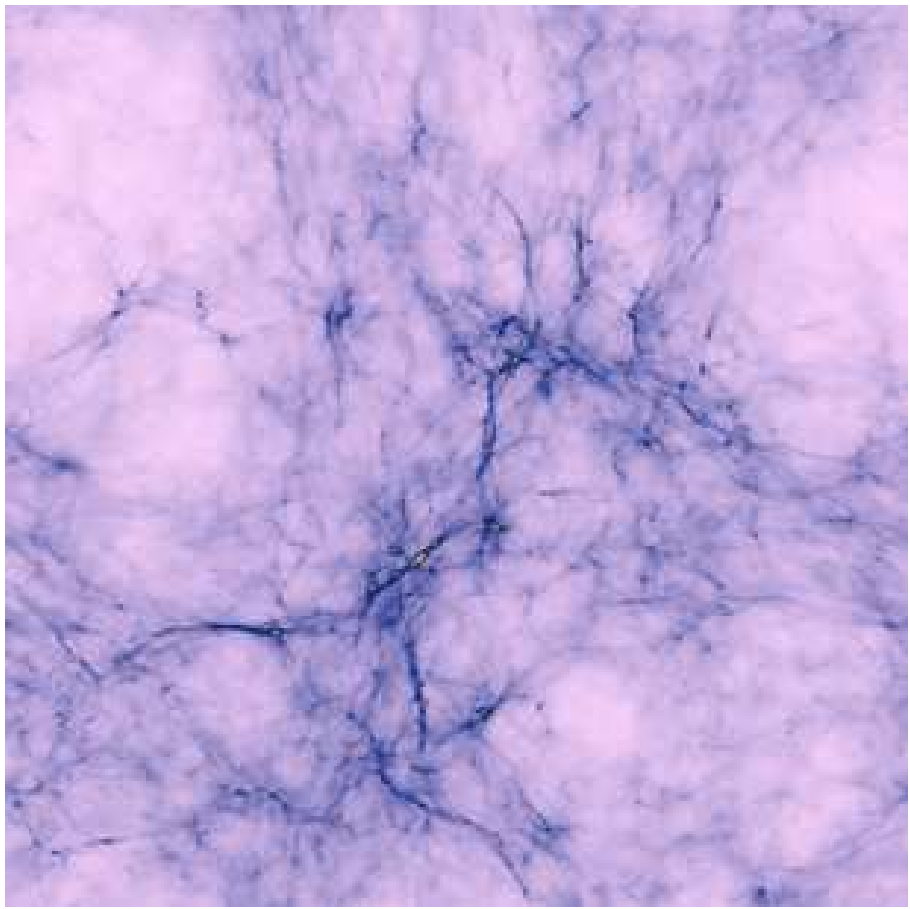}}
  \caption{The projected gas distribution at $z=17$ in a cubic volume of 600$h^{-1}$kpc 
on a side. The cooled dense gas clouds appear as bright spots at the intersections 
of the filamentary structures. From Yoshida et al. (2003)\cite{rf:YoshidaFirst}}
\label{fig:first}
\end{figure}

\subsection{Formation of the first cosmological objects\label{sec:early}}

Observations of distant quasars revealed that the intergalactic medium
was highly ionized at $z<6$. High resolution spectroscopic studies
of quasars at $z>6$ showed a significant increase in Ly-$\alpha$ absorption 
with the first detection of f Gunn-Peterson trough.\cite{rf:Fan,rf:Becker}
These results indicated that cosmic reionization completed at $z\sim 6-7$. 
However, results from the WMAP 
satellite have challenged this scenario by suggesting that
that the IGM is significantly ionized much earlier than inferred 
from the quasar observations.\cite{rf:Kogut} 
The Thomson optical depth determined from the temperature-polarization 
correlation is $\tau \sim 0.17$, suggesting that a large fraction of the IGM
was ionized at $z \sim 17$. While there is still a substantial 
uncertainty in this measurement, it is clear that
the first cosmic structure emerged very early on. 

Planned observational programs will exploit future instruments such as
{\it JWST}\footnote{http://ngst.gsfc.nasa.gov/} and 
{\it ALMA}\footnote{http://www.nro.nao.ac.jp/~lmsa/}
to probe the physical processes which shaped
the high-redshift Universe.  Among the relevant scientific issues are
the star formation rate at high redshift, the epoch of
reionization, and the fate of high-redshift systems.  
The statistical properties of early
baryonic objects are of direct relevance to understanding the
significance of the first stars to these phenomena.  In this context,
the key theoretical questions can be summarized as {\it when and where
did a large population of the first stars form?} and {\it how and when
did the Universe make the transition from primordial to ``ordinary''
star formation?}


The study of the cooling of primordial gas in the early universe 
and the origin of the first
baryonic objects has a long history\cite{rf:Matsuda,rf:Kashlinsky,
rf:Couchman} (see also the contribution by Haiman in this volume).
Recent numerical studies of the formation of primordial gas clouds
and the first stars indicate that this process likely began as early
as $z\approx 30$ in the CDM model.\cite{rf:ABN,rf:BCL,rf:YoshidaFirst}
In these simulations, dense, cold clouds of self-gravitating
molecular gas develop in the inner regions of small halos and contract
into proto-stellar objects with masses in the range $\approx 100 - 1000
M_{\odot}$. Fig.~\ref{fig:first} shows the projected gas distribution 
in a cosmological simulation that includes hydrodynamics and detailed 
chemistry.\cite{rf:YoshidaFirst}
The primordial star-forming gas clouds are found at
the nots of filaments, resembling galaxy clusters and filamentary
structure, although being much smaller in mass and size. 
(This manifest the hierarchical nature of structure in the CDM universe.)
The simulation evolved the non-equilibrium rate equations for 9 chemical 
species of primordial composition and include the relevant gas heating and cooling 
in a self-consistent manner.  From a large sample of dark halos, necessary conditions 
under which the first baryonic objects form are identified. 
 
\begin{figure}
  \centerline{\includegraphics[width=8cm]{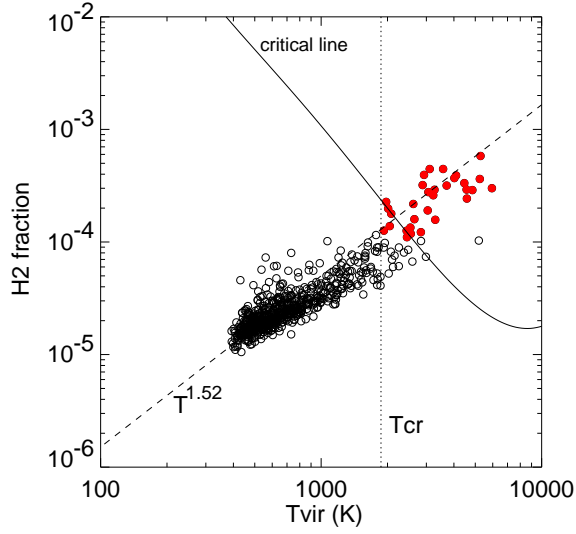}}
  \caption{The mass weighted mean H$_{2}$ fraction versus virial temperature 
for the halos that host gas clouds (filled circles) and for those that do not 
(open circles) at $z=17$. 
The solid curve is the H$_{2}$ fraction needed to cool the gas
at a given temperature and the dashed line is the asymptotic H$_{2}$
fraction.}
  \label{fig:first2}
\end{figure}

Fig.~\ref{fig:first2} shows the molecular fraction $f_{\rm H_{2}}$ 
against the virial temperature for halos. 

\begin{figure}
  \centerline{
    \includegraphics[width=7cm]{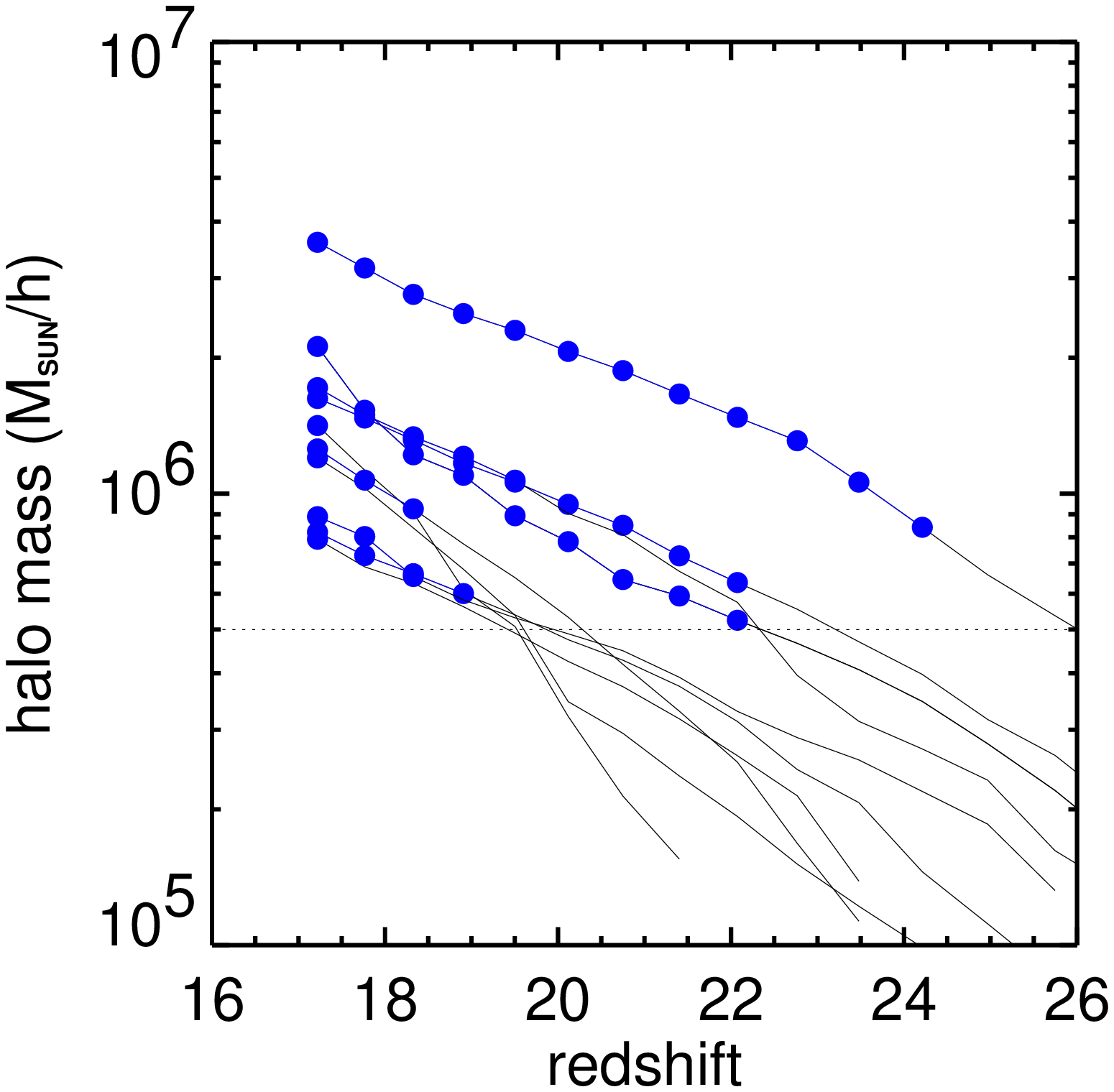}
    \includegraphics[width=7cm]{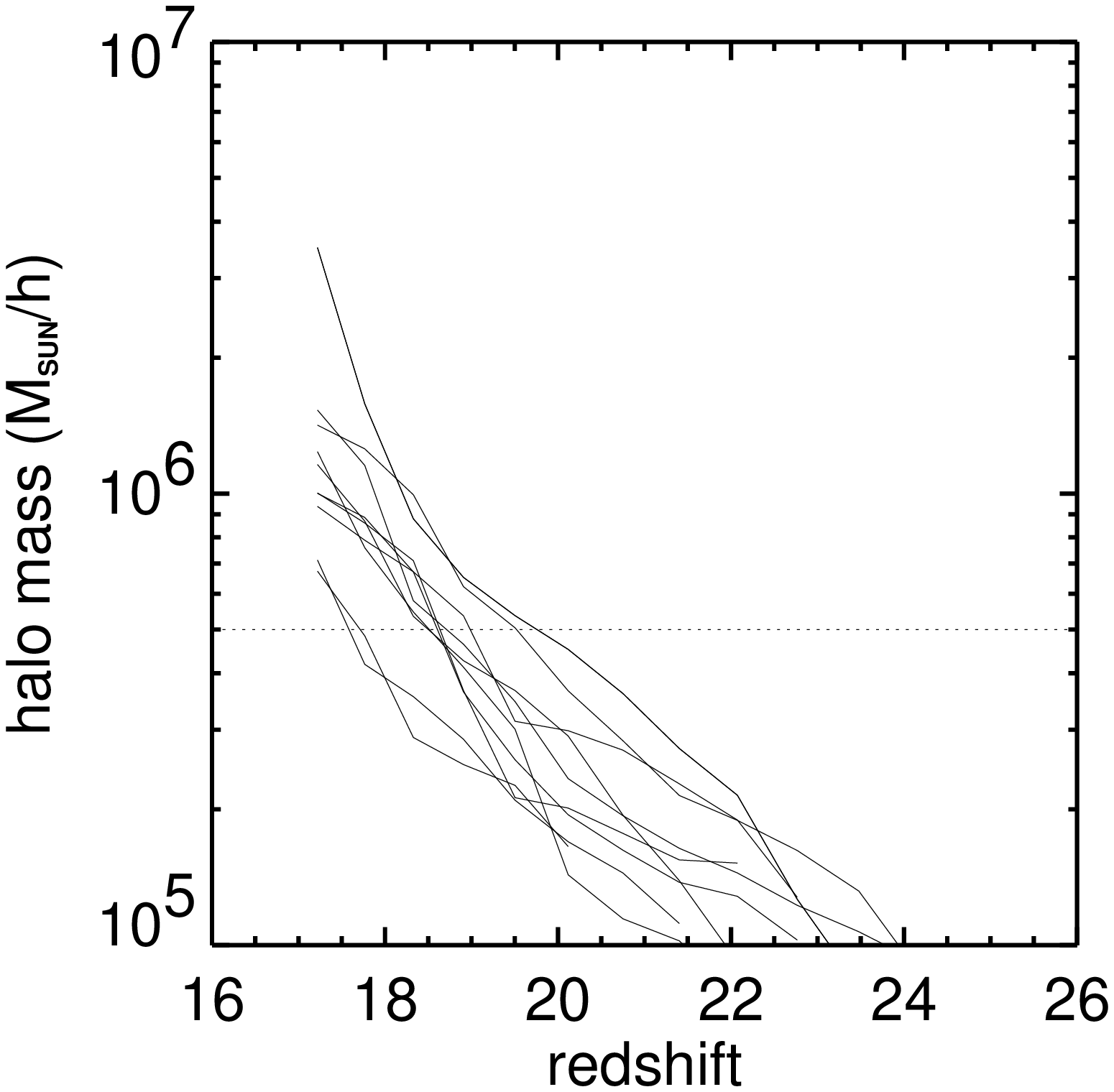}
  }
  \caption{The mass evolution of the halos that host gas clouds at z=17 (left)
and those that do not (right).}
  \label{fig:first3}
\end{figure}

\noindent
The solid line is an analytical estimate of the H$_{2}$ fraction needed to
cool the gas, which we compute {\it a l\`{a}} Tegmark et al.\cite{rf:Tegmark97}
In Fig.~\ref{fig:first2}, halos appear to be clearly separated into two
populations; those in which the gas has cooled (top-right), and the
others (bottom-left).  
The analytic estimate indeed agrees very well
with the distribution of gas in the $f_{\rm H_{2}}$ - $T$ plane.
However, it can be also seen that not only the H$_{2}$ fraction
determine whether or not the gas in halos can cool. 
In Fig.~\ref{fig:first3}, the mass evolution is plotted for a
subset of halos that host gas clouds (top-left panel) and of another
subset of halos that do not host gas clouds (top-right panel).  In the
top-left panel, filled circles indicate when they host
gas clouds. The figures show a clear difference between the two
subgroups in their mass evolution. Most of the halos in the top-left
panel experience a gradual mass increase since the time their masses
exceeded $M_{\rm cr}$, whereas those plotted in the top-right panel
have grown rapidly after $z\sim20$. It appears that the gas in halos
that accrete mass rapidly (primarily due to mergers) is unable to cool
efficiently. Therefore, ``minimum collapse mass''
models are a poor characterization of primordial gas cooling and gas
cloud formation, because these processes are significantly affected 
by the dynamics of gravitational collapse. This suggests that
it is important to take into account the details of halo formation 
history in the CDM model.

\subsection{Feedback from the first objects}
The birth and death of the first generation of stars have important
implications for the thermal state and chemical properties of the
intergalactic medium in the early universe.
The initially neutral, chemically pristine gas was
reionized by ultraviolet photons emitted from the first stars, but also
enriched with heavy elements when these stars ended their lives as
energetic supernovae.
The importance of supernova explosions, for instance, can be easily 
appreciated by noting that only light elements were produced during the 
nucleosynthesis phase in the early universe.  
Chemical elements heavier than lithium are thus thought to be
produced exclusively through stellar nucleosynthesis, and they must have
been expelled by supernovae to account for various observations of
high-redshift systems.  The destruction of star-forming regions 
by radiation from the first stars and/or by supernova explosions is also 
of considerable cosmological interest.  If the primordial gas cloud and 
the halo gas are completely blown away by a single supernova explosion, 
star-formation is quenched for a long time in the same region.


\subsubsection{Radiative feedback \label{sec:HII}}

The first feedback effect we consider is radiation from the first stars.
Cosmic reionization by stellar sources proceeds first by the formation 
of individual \HII regions around radiation sources (stars/galaxies), 
and then by percolation of the growing \HII bubbles
\cite{rf:Gnedin97,rf:Ricotti02,rf:Sokasian04}.
The shape and the extension of the individual \HII regions critically
determine the global topology of the ionized regions in a cosmological
volume at different epochs during reionization.

Studies on the formation of \HII regions in dense gas clouds date back to
the seminal work by Str\"omgren.\cite{rf:Stromgren} \ Since then the 
structure of \HII regions and the interaction with the surrounding medium 
have been extensively studied.\cite{rf:Yorke,rf:Franco} 
Recently, two groups carried out radiation hydrodynamics simulations 
of ionization front propagation around the first
stars.\cite{rf:Whalen,rf:Kitayama04}
The simulations start from a realistic initial density profile for 
primordial star-forming clouds and include gravitational forces exerted by the host
dark matter halo. These two conditions make the evolution different
from that of present-day local \HII regions.

\begin{figure}
  \centerline{\includegraphics[width=14cm]{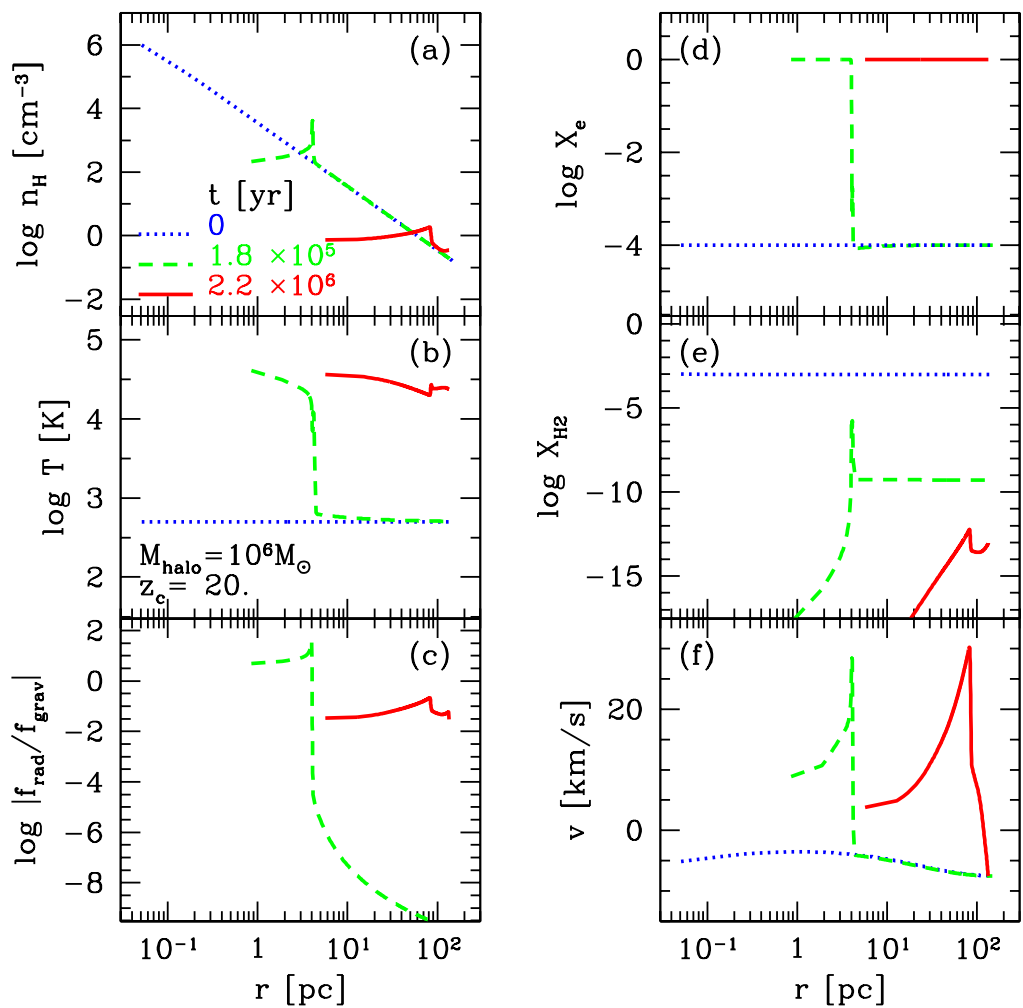}}
  \caption{Structure of an \HII region around a massive Population III star
inside a minihalo. Radial profiles of (a) hydrogen number density, (b) gas temperature,
(c) ratio of radiation force to gravitational force, (d) ionization fraction,
(e) molecular hydrogen fraction, and (f) radial velocity, at indicated output times
are shown. From Kitayama, Yoshida, Susa \& Umemura (2004)\cite{rf:Kitayama04}}
  \label{fig:HII}
\end{figure}

\begin{figure}
  \centerline{\includegraphics[width=11cm]{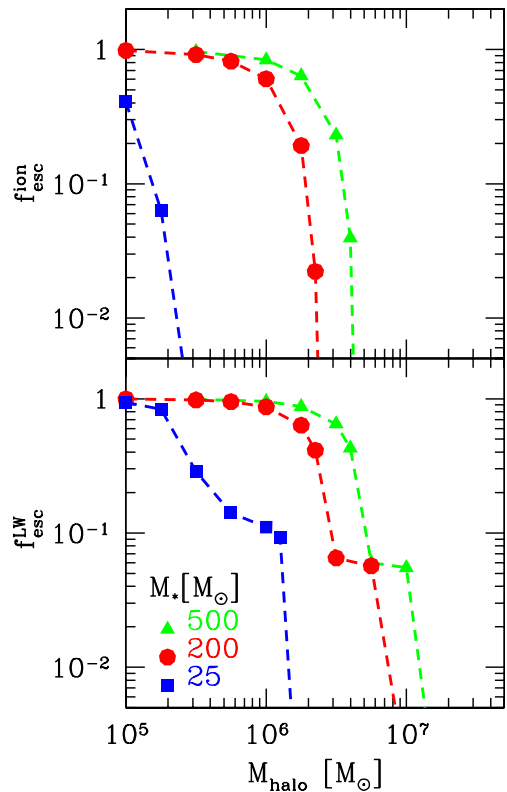}}
  \caption{Escape fractions of ionizing photons ($>13.6$eV)
and the Lyman-Werner photons (11.2-13.6eV) as a function of host halo mass.}
  \label{fig:HIIesc}
\end{figure}

\begin{figure}
  \centerline{\includegraphics[width=7cm]{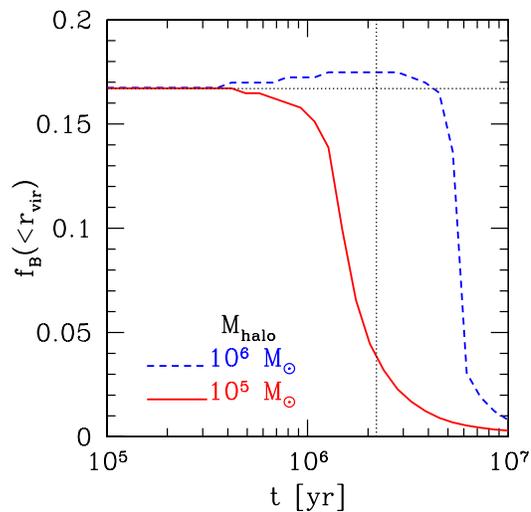}}
  \caption{Evolution of baryon fraction within virial radius in our fiducial runs.
The stellar radiation evacuates nearly all the gas is within 
a few to ten million years.\cite{rf:Kitayama04}}
  \label{fig:HIIevac}
\end{figure}

 Fig.~\ref{fig:HII} shows the radial profiles of various quantities
in an early \HII region.\cite{rf:Kitayama04}
The star-forming region is defined as a spherical dense molecular gas cloud 
with a power-law density profile within a dark matter halo, and a single 
massive Population III star with $M_{*}=200 M_{\odot}$ is embedded at the center.  
The formation of the \HII region is characterized by initial slow expansion 
of weak D-type ionization
front near the center, followed by rapid propagation of R-type front
throughout the outer gas envelope.  
The transition between the two front types 
is indeed a critical condition for the complete ionization of halos of 
cosmological interest.  In small mass ($< 10^6 M_{\odot}$) halos, the 
transition takes place within a few $10^5$ years, yielding high escape 
fractions ($>80\%$) of both ionizing and photodissociating photons.
Fig.~\ref{fig:HIIesc} shows the escape fractions for a large range
of halo mass. The gas in small mass halos is
effectively evacuated by a supersonic shock (see Fig.~\ref{fig:HIIevac}), 
with the mean density
within the halo decreasing to $\simlt 1 {\rm cm}^{-3}$ in a few million years. 
In larger mass ($> 10^7 M_{\odot}$) halos, on the other hand, the ionization 
front remains to be of 
D-type over the lifetime of the massive star, the \HII 
region is confined well inside the virial radius, and the resulting escape 
fractions are essentially zero.

The strong dependence of the photon escape fraction indicates
that only the first stars formed in small mass halos ($\simlt 10^6 M_{\odot}$)
can contribute to IGM ionization. It is also interesting that the 
critical mass for the escape of the Lyman-Werner photons, 
which dissociate hydrogen molecules, is slightly larger than that 
of ionizing photons. Nearly all the Lyman-Werner photons can escape
from halos with $M\sim 10^6 M_{\odot}$. 
A strong negative feedback is expected to be caused by these systems.

\subsubsection{Mechanical feedback}

Recent theoretical studies on the formation of primordial stars
consistently suggest that the first stars were rather massive
\cite{rf:ABN,rf:BCL,rf:OP03}, with an important exception
that those formed via filamentary collapse may be as small as
$\sim 1 M_{\odot}$.\cite{rf:Nakamura}  
If the first stars are indeed as massive as $\sim 200 M_{\odot}$, they
end their lives as energetic SNe via the pair-instability mechanism,
\cite{rf:Barkat,rf:Bond84,rf:HW02}
releasing a total energy of up to $\sim 10^{53}$~ergs.  
Such energetic 
explosions in the early universe are thought to be violently 
destructive: they expel the ambient gas out of the gravitational potential 
well of small-mass dark matter halos, causing an almost complete evacuation.
\cite{rf:BYH03,rf:Wada03} \ Since the massive stars process a substantial 
fraction of their mass into heavy
elements, early SN explosions may provide an efficient mechanism to
pollute the surrounding intergalactic medium.\cite{rf:YBH04}  

The  physics of  astrophysical blastwaves  has been  extensively studied
since early 70's.\cite{rf:Ikeuchi,rf:Bertschinger,rf:OstrikerMcKee}  
On a cosmological background, Ikeuchi\cite{rf:Ikeuchi} suggested
energetic   explosions  in   the   early  universe as  a large-scale
star-formation  and galaxy  formation mechanism. 
Population III supernova explosions  in the  early
universe were also considered as a trigger of star-formation.
\cite{rf:CBA84}
Modern numerical simulations have been carried out by two groups.
\cite{rf:BYH03,rf:Wada03} It has been shown that the expelled gas by supernovae
falls  back to the dark halo potential  well after about  the system's  
free-fall time.   While these
previous works consistently showed the {\it destructive} aspect of early
supernova  explosions,  they employed  idealized  or simplified  initial
conditions and hence the  precise effect remains uncertain.  The density
and  density  profile  around  the  supernova sites  are  of  particular
importance  because the  efficiency  of cooling  of  SNRs is  critically
determined by the density inside the blastwave.

As shown in the previous section,
for `mini-halos' with mass $\sim 10^6 M_{\odot}$, I-fronts 
quickly expand to a radius of over 1 kpc and the halo gas is effectively 
evacuated.  Interestingly, the final gas distribution is very different 
for cases with small ($\sim 10^6 M_{\odot}$) and large 
($\sim 10^7 M_{\odot}$) mass halos.  The radial profiles of density, 
temperature and velocity at the death of the central star should provide 
appropriate initial conditions for the studies of subsequent SN feedback.

\begin{figure}
  \centerline{\includegraphics[width=14cm]{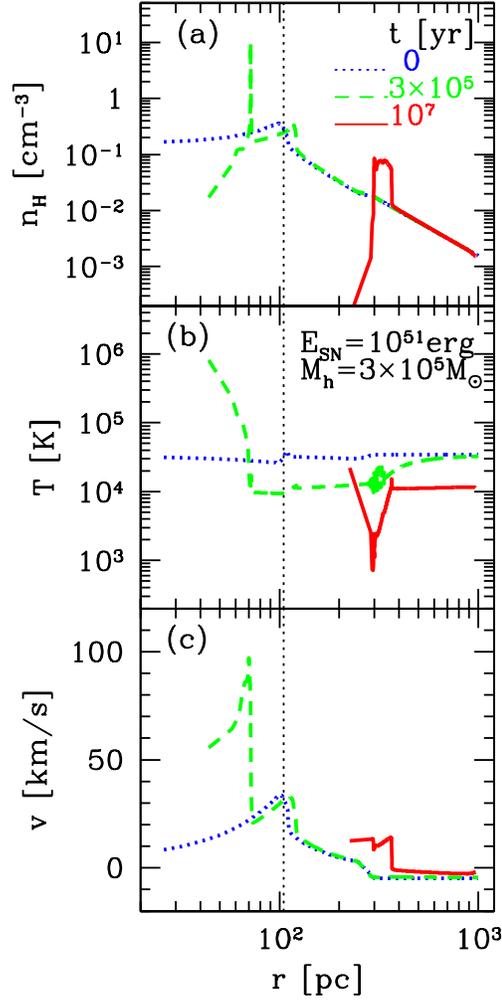}}
  \caption{Evolution of the early SNR in the case of
$E_{\rm SN}=10^{51}$ erg $M_{\rm host}= 3.2 \times 10^5$ M$_\odot$, and
$M_{\rm s}=200$ M$_\odot$; (a) hydrogen density, (b) gas temperature,
and (c) outward velocity at $t=0$ (dotted lines), $3\times 10^5$
(dashed) and $10^7$ yr (solid), where $t$ denotes the time elapsed since
the end of the free expansion stage.  The vertical dotted line indicates
the virial radius of the host halo. From Kitayama and Yoshida (2005).\cite{rf:Kitayama05}}
\label{fig:snprof}
\end{figure}

Kitayama and Yoshida\cite{rf:Kitayama05}
carried out hydrodynamic simulations of radiative supernovae remnants at $z\sim 20$,
starting the simulations from the resulting density profiles in the first \HII
regions. Fig.~\ref{fig:snprof} shows the evolution of the radial profiles of 
various quantities around the
supernova site. The blastwave quickly propagates over the halo's virial
radius, leading to complete evacuation of the gas even
with the input energy of $10^{51}$ erg. 
A large fraction of the remnant's thermal energy is lost in $10^5-10^7$ yr 
by line cooling, whereas, for larger explosion energies, the remnant cools mainly 
via inverse Compton scattering.
In the early universe, the inverse Compton process with cosmic background photons
acts as an efficient cooling process. 

The situation drastically changes if there were no I-front expansion
prior to the SN explosion. Such cases may be realized when ionizing photons
do not break out from the very central region and 
a ultra-compact \HII region is formed. 
If the initial density profile of the run
shown in Fig.~\ref{fig:snprof} is modified to a pure power-law with
$\rho \propto r^{-2}$, assuming that the density profile has not been
significantly modified by radiation, the cooling time of the inner-most shell gets
extremely small and the ejected energy is rapidly lost during
the free-expansion stage.
Accordingly the blastwave {\it stalls} in the dense environment 
and the halo gas in the outer envelope will not be
undisturbed.  This is the case even if the ejected SN energy is much
greater than the binding energy for baryons within the virial radius.
This clearly shows the importance
of setting-up appropriate initial configurations in quantifying the
degree of SN feedback and that simple analytic estimates based on
explosion energy to binding energy ratio are unreliable.

\begin{figure}
  \centerline{\includegraphics[width=10cm]{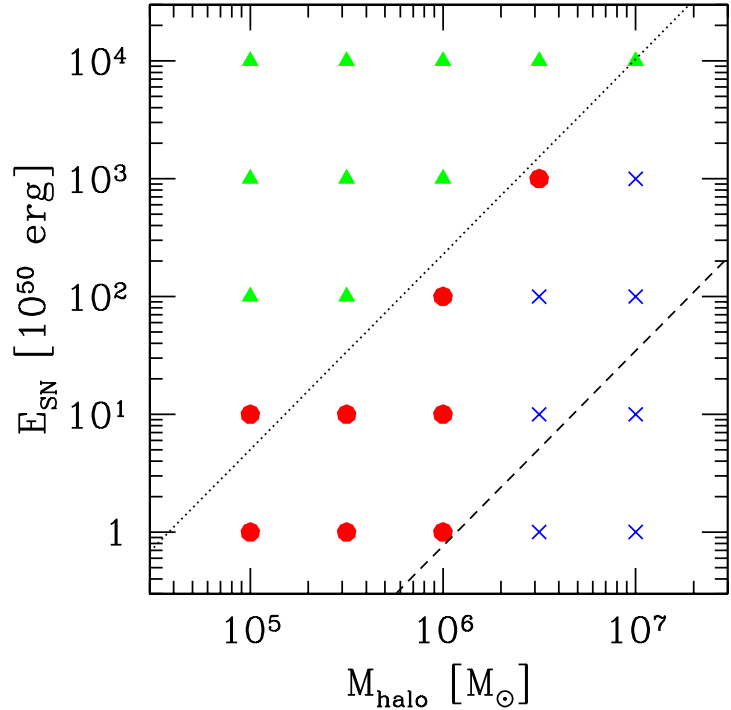}}
  \caption{Destruction efficiency of the first supernovae. 
Halos blown away even in the absence of initial
I-front expansion are marked by triangles, those blown away only in the
presence of I-front initial expansion by circles, and those not blown
away by crosses.  Dotted and dashed lines show the binding energy of the
gas for a given $M_{\rm h}$ and 300 times the same quantity,
respectively. }
\label{fig:sn}
\end{figure}

\noindent
Fig.~\ref{fig:sn} summarizes the results from a series of calculations
of Kitayama \& Yoshida.\cite{rf:Kitayama05}
A simple criterion, $E_{\rm SN} > E_{\rm bi}$, where 
$E_{\rm bi}$ is the gravitational 
binding energy, is often used to determine the destruction efficiency.
However, whether or not the halo gas is effectively blow-away is determined not only
by the host halo mass (which gives an estimate of $E_{\rm bi}$), but also
by a complex interplay of hydrodynamics and radiative processes.
SNRs in dense environments are highly radiative and thus
a large fraction of the explosion energy can be quickly radiated away.
An immediate implication from the result is that, in order for the processed 
metals to be transported out of the halo and distributed to the IGM,
I-front propagation and pre-evacuation of the gas must precede the
supernova explosion. This roughly limits the mass of host halos from which 
metals can be ejected to $<10^6 M_{\odot}$ for explosion energy $\simlt 10^{53}$
erg.

As has been often discussed in the literature\cite{rf:Wada03,rf:YBH04}, 
the feedback effects from the first stars tend to quench further
star-formation in the same place.  Although metal-enrichment by the
first supernovae could greatly enhance the gas cooling efficiency, which
would then change the mode of star-formation to that dominated by
low-mass stars,\cite{rf:BL03}
the onset of this `second-generation' stars may be delayed particularly
in low-mass halos. Hence early star-formation is self-regulating;
if the first stars are massive, only one period of star-formation is 
possible for a small halo and its descendants within a Hubble time then.
The sharp decline in the efficiency of both feedback effects at 
$M_{\rm halo} >10^7 M_{\odot}$ (see Fig.~\ref{fig:HIIesc}, \ref{fig:sn})
indicates that the global cosmic star formation activity increases only 
after such larger halos start forming. 
An important question remains, however.
Without metal-enrichment, gas cooling efficiency is still limited
even though hydrogen atomic line cooling becomes effective in halos
with $T_{\rm vir} > 10^4$K. Oh \& Haiman\cite{rf:OhHaiman} argue 
that H$_{2}$ molecules are needed as main coolants
for the primordial gas to further fragments in a dense gaseous disk
in large halos. If the formation of H$_{2}$ is strongly suppressed
by a soft-UV background, star-formation does not take place,
or the first black holes may be formed in low-spin halos.\cite{rf:BLBH}
Further studies on star-formation in such large halos are 
clearly needed.

\begin{figure}
  \centerline{\includegraphics[width=13.5cm]{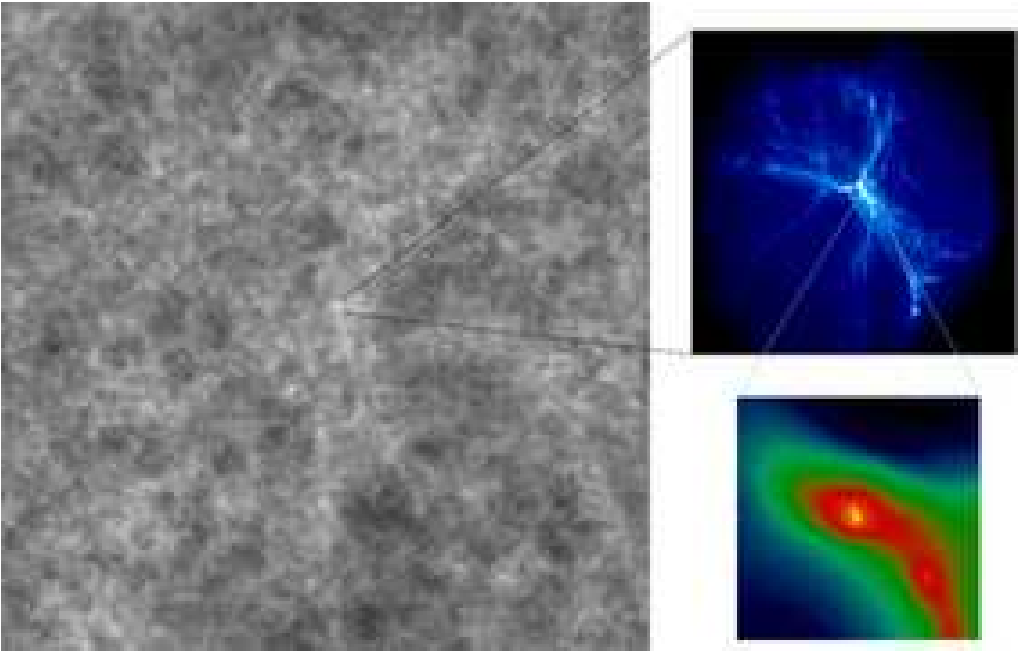}}
  \caption{The large-scale parent simulation at $z=20$
showing the highest density peak within a portion of 100 $h^{-1}$Mpc 
on a side (left panel). The earliest object formed in this peak 
region at $z=49$ is shown on the right.
The bottom-right panel shows the projected gas density field in a 
cubic volume of 100 pc on a side centered at the primordial gas cloud.
From Gao et al. (2005)\cite{rf:Gao05}}
  \label{fig:gao}
\end{figure}

\subsection{The earliest object in a $\Lambda$CDM universe}

Numerical simulations and analytic models 
of the first cosmic structure to date mostly addressed
the formation of the first objects within a hypothetical volume 
 or in a statistical sense.
As has been discussed in the previous sections, 
halos with $T_{\rm vir}\sim 1000-2000$ K 
are usually thought to be hosts of the first stars;
those collapsing at $z\sim 20$ from 3-$\sigma$ density peaks 
typically correspond to these.
However, nonlinear growth of the high density peaks 
and the formation of the first objects could occur
in a significantly biased manner. 
They could also cause large-scale feedback effects.
Hence any statistical argument based on averaged quantities
in a `mean' density universe 
may miss important aspects of early structure formation.
An interesting question in this context is, {\it when and where did the very 
first star form ?}

An analytic estimate based on the Extended Press-Schechter
theory tells that the highest redshift object
in the observable volume of an observer at $z=0$ 
should be one formed from an 8-$\sigma$ 
fluctuation at $z\sim 48$.\cite{rf:Miralda} 
In the CDM model, the earliest structures are expected to be
such extremely rare objects, and thus it is very unlikely
to be found in usual cosmological simulations employing a small volume.
Recently Gao et al.\cite{rf:Gao05} carried out a 
series of very high resolution re-simulations 
to locate one of the earliest objects in a {\it cosmological} volume.
The multi-level re-simulations were set-up
and carried out in the following manner. A very massive halo is identified in 
a large simulation box of (0.68Gpc)$^3$ volume at $z=0$. 
This cluster and its immediate surroundings are re-simulated with a 
higher mass resolution.
Then a halo merging tree is constructed and 
the progenitor, which contributes most
in mass, is identified at some earlier epoch. The progenitor is again re-simulated
at a higher mass resolution using a zoom-in technique. 
At each refinement level, small length-scale perturbations are added
to the particle displacement field in order to realize a proper
initial condition for the given CDM power spectrum.
In their highest resolution simulation, the gas particle mass
is $0.34 h^{-1}M_{\odot}$ and that of dark matter is 
$2.2 h^{-1}M_{\odot}$. 
Fig.~\ref{fig:gao} shows a portion of the parent simulation 
and the earliest object in the highest resolution simulation.
Significant small-scale clustering is seen already at $z=49$.
A halo with mass $3\times 10^5 M_{\odot}$ is formed at the center,
and the gas within it has reached the characteristic state 
of the primordial molecular gas cloud with
temperature $T\sim 200$ K and particle number density $n_{\rm H}\sim 10^4 
{\rm cm}^{-3}$.
The Jeans mass for these values is $M_{\rm jeans}\sim 3000 M_{\odot}$, and 
the cloud mass has already exceeded  $M_{\rm jeans}$ at $z=49$.
The gas cloud is thus expected to collapse, although it occurs slowly.

Since these earliest objects are extremely rare, 
the net feedback effect from them is limited to
its surroundings. For instance, even if a very massive star
is formed in the gas cloud, the Lyman-Werner photons
would just escape from the region and absorbed by relic intergalactic H$_2$.
Within the same halo, the subsequent evolution of the gas 
and the formation of the second-generation objects
may be significantly 
disturbed by them, via various effects as discussed in 
the previous sections. In particular, if very massive stars are
formed in these objects and end their lives as energetic supernova,
they expel heavy elements such as carbon and oxygen,
hence dramatically changing the gas cooling efficiency.
\cite{rf:BL03} 
In such cases, the gas is first expelled from the host
halo, which has just a mass of $3\times 10^5 M_{\odot}$ at z=49,
and eventually falls back when the halo grows 
and becomes more massive. This fall-back time scale is
crudely estimated to be {\it at least} of the order of 
the system's dynamical time, which is $\sim 10$ Myrs at $z=49$. 
This could be much delayed 
because the expelled gas is initially moving outward with a
velocity greater than the halo's virial velocity 
(see the bottom panel in Fig.~\ref{fig:snprof}).
It is found that the host halo mass of the `earliest' object in the simulation
increases to $M=2\times 10^6 h^{-1}M_{\odot}$ by $z=35$.\cite{rf:Gao05}
The corresponding virial temperature is just above $10^4$K,
at which the gas can cool by hydrogen atomic transitions.
Hence the formation of the second-generation object,
the first proto-galaxy, is expected to occur in the same region
only at $z\simlt 35$. This epoch is still earlier than the collapse
epoch of $10^6 M_{\odot}$ halos from 3-$\sigma$ density peaks. 
If stars are formed efficiently in these pre-galactic objects,
they will be the brightest sources at the epoch and may contribute 
significantly to reionization.

\subsection{Prospects for observations}
We have discussed the physics of early structure formation 
and presented recent development in theoretical studies.
We would like to close this chapter by mentioning future observations. 
The nature of the sources of the first light and the origin of heavy elements 
have not been determined yet, but 
the prospects for observationally revealing these issues
in the near future appear bright.
The ongoing operation of {\it WMAP} will yield a more precise
value for the total optical depth to reionization.
In the longer term, post-{\it WMAP} CMB polarization experiments such
as {\it Planck} will probe the reionization history.\cite{rf:Kap03} 
Detection of small-angular scale
CMB fluctuations and, particularly, the second-order
polarization anisotropies on arcminute scales can place
strong constraint on the details of reionization.
\cite{rf:Vish87,rf:Liu01,rf:MSantos03}
Near-infrared observations of afterglows from high-redshift
gamma-ray bursts can also be used to probe the reionization history
at possibly $z>10$.\cite{rf:BarL04,rf:InoueAK03,rf:Sinoue03,rf:Ioka}
Mapping the morphological evolution
of reionization may be possible by observations of
redshifted 21cm emission.\cite{rf:ScottRees,rf:Bagla}
in particular by the Square Kilometer Array
and LOFAR.\cite{rf:Tozzi,rf:Iliev03,rf:Furl04}
Analyses of these various high-precision data promise to provide a more 
complete picture of cosmic reionization and possibly the matter density 
distribution in the early Universe over a wide
range of scales and its relationship to the formation of stars and
galaxies.
The precise measurement of the near-IR cosmic background radiation
will constrain the total amount of light from early generation
stars.\cite{rf:Bond86,rf:Santos,rf:Cooray04}
Ultimately, direct imaging and spectroscopic observations of high redshift
star clusters by the {\it James Webb Space Telescope} will
probe the evolution of stellar populations up to $z\sim 10-15$.
\cite{rf:OhRees01,rf:Jason,rf:Stia04}

Measurements of the relative abundances of
various heavy elements in metal-poor stars
should provide valuable information on the formation 
history of our Milky Way as well as the chemical evolution
of the universe.\cite{rf:Burris}
Interestingly, a strong argument against
very massive ($>140 M_{\odot}$) stars comes from
the observed abundance pattern of C-rich, extremely Fe-deficient
stars.\cite{rf:Christ02,rf:Sch03}
It remains to be seen whether or not such stars are
truly second generation stars and their elemental abundances
should precisely reflect the metal-yield from
the first supernovae. Observations of a large number of extremely
metal-poor stars will construct better statistics
\cite{rf:Frebel05} and improve
constraints on any models for the early chemical evolution.
Understanding the origin of the first heavy elements
in the universe and the nature of the sources
that are responsible for cosmic reionization will require the concerted use of
data from these broad classes of observations.

\section{Non-standard models: alternative to CDM ?}

The cold dark matter model has become the leading theoretical framework 
for the formation of structure in the Universe. 
While a broad range of recent observations provided strong support
for the $\Lambda$CDM model in which cold dark matter and dark energy dominate,
the lack of experimental evidence of such dark components 
and the fact that the physical origin and nature of them remain unknown
make the model still speculative. Thus it appears to be worth exploring
alternative scenarios and consider structure formation in non-standard
models.

 Modified Newtonian Dynamics (MOND) has been often 
proposed as an alternative to dark matter in many different contexts,
from the internal dynamics of galaxies\cite{rf:Milgrom}
to large-scale structure.\cite{rf:Saunders}
While so far a number of arguments against MOND have been made 
based on various observations,\cite{rf:Scott,rf:Aguirre,rf:McGaugh} 
MOND appears to die hard. Moreover, it has had its own intrinsic problem
that the corresponding relativistic theory does not exist.
Bekenstein\cite{rf:Bekenstein04} recently made the first attempt 
to construct a relativistic MOND theory.
It will be of considerable interest if the formation and evolution of 
structure in a MONDian universe can be addressed in a fully 
self-consistent manner.

More recently, several specific attempts were made to construct self-consistent 
cosmological models including the deviation from Newton's law on cosmological 
scales. For instance, Dvali, Gabadadze and Porrati\cite{rf:Dvali} proposed a 
scenario to explain the accelerating universe as a result of leaking gravity to 
extra dimension in the context of braneworld model.  
According to this model, the accelerating universe can be accounted for without 
dark energy component, but due to the modification of Newton's law of gravity on
cosmological scales. Other models which suggest deviations from Newton gravity 
on cosmological scales include a ghost condensation model
\cite{rf:Arkani} and scalar-tensor theories \cite{rf:Acquaviva}.
A few attempts have been made to constrain deviations from the
inverse-square Newtonian law of gravity on cosmological scales
\cite{rf:Sealfon,rf:Nusser,rf:Shirata}. 
Shirata et al.\cite{rf:Shirata} 
consider specifically deviation 
that is described by an additional Yukawa-like term.
They derived constraints on the amplitude and the length scale
by comparing the predicted matter 
power spectra in this non-Newtonian model 
and the galaxy-galaxy power spectrum from the SDSS, on the
assumption that galaxies are linearly biased tracer of the
underlying mass. As shown by Shirata et al., 
large-scale structure as probed by galaxy
clustering or gravitational lensing observations may 
provide a unique tool to test the Newtonian gravity
at cosmological length scales $\gg 1$Mpc.

Another context in which CDM models are often claimed 
to be in trouble is the `fine' structure of dark matter halos,
typically that of galactic-size halos.
Despite a great amount of works devoted to the issue
of ``core or cusp'' in the past several years in both theory and observations,
there hasn't been any clear resolution nor even a consensus 
on whether or not the problem is real rather than apparent.
From theoretical interest, relatively minor modifications to the CDM 
model have been proposed. Such models invoke additional properties of 
dark matter particles \cite{rf:SIDM,rf:YosSIDM,rf:Bode01} or slight 
change in the primordial power spectrum\cite{rf:Kamion}. These
models, however, either have their own difficulties or lack strong motivation
from fundamental physics, and thus they are now thought to be 
unattractive alternatives. 

The angular resolution of observations 
(and other technical details) of rotation curve measurements 
are often ascribed as the source of the discrepancy.
However, even the most recent high resolution H-$\alpha$ observations
still show some convincing cases where the central density
profiles are rather flattened, being in conflict with
the CDM prediction.\cite{rf:JSimon}
Interpreting the measured rotation curves is another complex issue.
It has been suggested that reconstructing the
density profile of a triaxial dark halo 
from rotation curve measurements is nontrivial.
For some viewing angles the density profiles can appear
much shallower than the actual profile.\cite{rf:Hayashi}
It seems that the key point in this issue is to conduct 
a fair comparison between observation and simulations.
Detailed studies on galaxy and gas kinematics in CDM halos
will eventually provide a resolution to the discrepancy.

\section{Conclusions}
We would like to summarize this article by making
remarks to the following three important questions 
in cosmology:
\\

\begin{enumerate}

\item {\it How are luminous matter and dark matter distributed
in the universe, and what is the origin of bias ?}\\
Current generation galaxy redshift surveys are providing
a detailed picture of galaxy distribution in the local
universe. It will soon become possible to probe the distribution
of dark matter by weak gravitational lensing observations,
and that of diffuse baryons using X-ray telescopes. 
By making a complete `map' of all these components, we will 
discover {\it differences} in their distributions, and 
will obtain a comprehensive knowledge on the process of structure formation.
\\

\item {\it How and when did the first stars and the first galaxies form ?}\\
Promisingly, a number of observational plans are underway, to detect
CMB polarizations, metal-poor relic stars, signatures in infrared background, 
metals in high-$z$ Lyman-$\alpha$ forests, and faint light from very high-$z$ galaxies.
Understanding the origin of the first heavy elements
in the universe and the nature of the sources 
that are responsible for cosmic reionization will require the concerted use of
data from these broad classes of observations. 
To this end, theoretical (as opposed to phenomenological) astrophysics can
play a role.
\\

\item {\it We all bet for the $\Lambda$CDM model ?}\\
While there appear to be some conflictions
with observations and possible theoretical difficulties such as 
the `unnatural' value of $\Lambda$, it may be fair to say that there 
is yet no strong case against the $\Lambda$ + Cold Dark Matter model.
Still the most fundamental issue remains; the physical origin and nature 
of the dark components. The direct laboratory detection 
of massive particles would provide the most convincing confirmation
of the dark matter paradigm. Strong motivations for the existence 
of dark energy from fundamental physics must be explored;
otherwise worth seeking alternatives perhaps in the theory of gravity.

\end{enumerate}

\section{Acknowledgment}
NY thanks Nishinomiya city and the organizers of Nishinomiya-Yukawa Memorial
Symposium. Some of the figures in this contribution are kindly provided by
Takashi Hamana, Kohji Yoshikawa, Tetsu Kitayama, and Liang Gao. 
Support from the 21$^{\rm st}$ 
COE Program ``The Origins of the Universe and Matter'' at Nagoya University 
is greatly acknowledged.

\end{document}